## Article

# Temperature-Dependent Functions of the Electron–Neutral Momentum Transfer Collision Cross Sections of Selected Combustion Plasma Species

Osama A. Marzouk

College of Engineering, University of Buraimi, Al Buraimi 512, Oman; osama.m@uob.edu.om

**Abstract:** The collision cross sections (CCS), momentum transfer cross sections (MTCS), or scattering cross sections (SCS) of an electron–neutral pair are important components for computing the electric conductivity of a plasma gas. Larger collision cross sections for electrons moving freely within neutral particles (molecules or atoms) cause more scattering of these electrons by the neutral particles, which leads to degraded electron mobility, and thus reduced electric conductivity of the plasma gas that consists of electrons, neutral particles, and ions. The present work aimed to identify the level of disagreement between four different methods for describing how electron–neutral collision cross sections vary when they are treated as a function of electron temperature alone. These four methods are based on data or models previously reported in the literature. The analysis covered six selected gaseous species that are relevant to combustion plasma, which are as follows: carbon monoxide (CO), carbon dioxide ($CO_2$), molecular hydrogen ($H_2$), water vapor ($H_2O$), potassium vapor (K), and molecular oxygen ($O_2$). The temperature dependence of the collision cross sections for these species was investigated in the range from 2000 K to 3000 K, which is suitable for both conventional air–fuel combustion and elevated-temperature oxygen–fuel (oxy-fuel) combustion. The findings of the present study suggest that linear functions are enough to describe the variations in the collision cross sections of the considered species in the temperature range of interest for combustion plasma. Also, the values of the coefficient of variation (defined as the sample standard deviation divided by the mean) in the collision cross sections using the four methods were approximately 27% for CO, 42% for $CO_2$, 13% for $H_2$, 39% for $H_2O$, 44% for K, and 19% for $O_2$. The information provided herein can assist in simulating magnetohydrodynamic (MHD) power generators using computational fluid dynamics (CFD) models for combustion plasma flows.

**Keywords:** collision cross section; CCS; momentum transfer cross section; MTCS; electron–neutral scattering; plasma; MHD; magnetohydrodynamic

## 1. Introduction

### 1.1. Background

In partially ionized plasma, electrons, neutral particles (molecules or atoms), and ions coexist. The collision cross sections (scattering cross sections or momentum transfer cross sections) of the interactions between the electrons and the neutral particles (the neutrals) of each gaseous species in the plasma are numbers that influence the mobility of these electrons within these neutrals. Rather than being geometric cross-sectional areas, they actually describe the chance of a collision event [1]. Higher collision cross sections result in more frequent electron–neutral collisions, which, in turn, result in a shorter mean free path of the electrons. This restriction on the continuous straight-line movement of the electrons causes a decline in the plasma electric conductivity [2].

For a single gaseous species "s" with a number density ($n_s$) and with a collision cross section ($Q_s$) for the particles of that species, the free path length ($\lambda_e$) of the electrons colliding with the neutral particles of that species is

$$\lambda_e = \frac{1}{n_s \, Q_s} \tag{1}$$

And the electron–neutral collision cyclic frequency is

$$\nu_{e-n}(v_e) = n_s \, Q_s(v_e) \, v_e \tag{2}$$

where ($v_e$) is the electron speed (velocity magnitude), which is amplified at higher temperatures, and which also can be induced via an applied magnetic field that the electron experiences when moving parallel to that magnetic field.

The time between two successive electron–neutral collisions is the reciprocal of the electron–neutral collision cyclic frequency. Thus, the time between electron–neutral collisions is expressed as

$$\tau_{e-n}(v_e) = \frac{1}{n_s \, Q_s(v_e) \, v_e} \tag{3}$$

If the average time between electron–neutral collisions is denoted as $\overline{\tau}_{e-n}$, with a corresponding average electron–neutral collision cyclic frequency of $\overline{\nu}_{e-n}$, and a corresponding average collision cross section of $\overline{Q}_s$, and also an average electron speed of $\overline{v}_e$, then these quantities are related as follows:

$$\overline{\tau}_{e-n} = \frac{1}{n_s \, \overline{Q}_s \, \overline{v}_e} \tag{4}$$

$$\overline{\nu}_{e-n} = n_s \, \overline{Q}_s \, \overline{v}_e \tag{5}$$

And the electron mobility can then be found as

$$\mu_{e-n} = \frac{e \, \overline{\tau}_{e-n}}{m_e} = \frac{e}{m_e \, \overline{\nu}_{e-n}} = \frac{e}{m_e \, n_s \, \overline{Q}_s \, \overline{v}_e} \tag{6}$$

And the electric conductivity of the plasma gas due to the movement of these mobile electrons is

$$\sigma_{e-n} = n_e \, e \, \mu_{e-n} = \frac{n_e \, e^2 \, \overline{\tau}_{e-n}}{m_e} = \frac{n_e \, e^2}{m_e \, \overline{\nu}_{e-n}} = \frac{n_e \, e^2}{m_e \, n_s \, \overline{Q}_s \, \overline{v}_e} \tag{7}$$

If it is important to maximize plasma electric conductivity, such as in magnetohydrodynamic (MHD) power generation [3–6], the electron collision cross sections should be minimized.

The topic of the collision cross sections of electrons has received research attention in several studies over many years [7–15]. The collision cross section of an electron depends on the kinetic energy of the electron and thus depends on its speed [16,17]. However, such a detailed description can be too cumbersome and time-consuming for practically estimating the electric conductivity of plasma, where finding the electron–neutral collision cross sections is just one stage of several stages of computation needed to find a single electric conductivity value for a plasma gas mixture. For example, finding the Coulomb scattering of electrons due to ions and other electrons, and finding the number densities of each neutral species, are other stages of computations that should be completed as prerequisites for estimating the electric conductivity [18,19].

Successful simulation of magnetohydrodynamic (MHD) problems, in order to analyze or to design MHD systems (in the form of a channel), in the application field of electric power generation may require applying the computational fluid dynamics (CFD) method to numerically solve a set of various partial differential equations that govern the spatial distribution and the temporal variations in density, pressure, velocity vector, temperature, electric potential, and electric current density vector for the plasma gas flowing through the MHD system [20–22]. In this field of MHD and plasma gases, ordinary combustion product gases that are electrically non-conductive (not ionized) can be made electrically conductive by adding a small amount of a gas (or a compound of it) that is partly ionized at typical combustion product temperatures without other sophisticated ionization methods. This additive ionizable substance (also called the seed) can be cesium vapor, potassium vapor, or a compound containing either of them. Some of the electrons in the seed atoms are liberated under the influence of the high temperature of the carrier combustion product gases, making the gaseous mixture electrically conductive, thus making it a plasma gas [23]. Knowing the chemical composition and the electric conductivity of this plasma gas is an important step during the CFD modeling, which eventually allows for predicting the electric power output that can be extracted from the flowing plasma gas inside the MHD system, which acts as a very powerful source of electricity. Therefore, there is a relationship between CFD, MHD, combustion, and plasma simulation.

In a CFD model for MHD problems, the spatial zone (the MHD channel) of high-speed (possibly supersonic speed) traveling plasma may be discretized into thousands or even millions of small finite volumes (called computational cells) or into a huge number of nodes [24–26], and the electric conductivity in each cell or node is computed as a plasma property, among other properties (such as thermal conductivity and viscosity). This is further repeated in order to advance the solution in time or to ultimately reach a steady condition, where it becomes possible to estimate how much electric power the MHD system can supply to a connected external load.

It is highly advantageous that the plasma electric conductivity can be calculated by knowing only the local temperature, the local pressure (which allows knowing the number density of particles or the mass density of their species), and the local chemical composition, because these are readily obtained variables during an ongoing CFD simulation. Thus, expressing the electron–neutral collision cross sections for each involved species as a simple function of the electron temperature avoids the need for additional complex computations in case these cross sections are further modeled as depending also on electron thermal speed (speed due to temperature), which is traditionally subject to the Maxwell–Boltzmann probability distribution [27,28]. Approximating the electron–neutral collision cross sections for a given chemical gaseous species as a function of temperature, as the only independent variable, greatly simplifies the process of computing the electric conductivity later. This simplification is common and has been adopted in different plasma studies, where a monoenergetic (electron-speed-dependent) distribution is replaced with a single constant (mean or effective) value [29–31].

*1.2. Goal and Benefits of This Study*

The present study considered four sources of published data on electron–neutral collision cross sections for scattering by six species, which represent five gaseous species common in combustion, either as combustion reactants (molecular hydrogen and molecular oxygen; $H_2$ and $O_2$) or as combustion products (carbon dioxide, carbon monoxide, and water vapor; CO, $CO_2$, and $H_2O$). The sixth species is potassium vapor, which is considered an important seeded element (added in a small amount) to facilitate ionization and thus to act as the main source of electrons in the plasma [32]. The electron–neutral collision cross sections were investigated for a range of absolute electron temperatures from 2000 K to 3000 K, which is a suitable range for representing conventional air–fuel combustion with unheated air, hotter-air combustion with preheated air, or even-hotter oxygen–fuel (oxy–fuel) combustion [33–36].

In the present study, (1) four different methods of electron–neutral collision cross sections for combustion plasma are presented. This can help plasma modeling researchers in comparing these methods and utilizing the appropriate one for their CFD models. The results are discussed, and connections with the electric conductivity and combustion type (air–fuel versus oxy–fuel) are established. (2) Quantitative measures of the disparity between the four methods of computing electron–neutral collision cross sections are reported, which indicate the reasonableness of applying a speed-independent approximation given the implied high levels of uncertainties. (3) Linear regression models for the electron–neutral collision cross sections are developed based on discrete values reported in one of the sources, for the six species covered herein. (4) The idea of implementing the mean kinetic energy in the monoenergetic collision cross section functions for the six species is tested, and it is found to give acceptable results in most of the covered species. (5) Examples of the monoenergetic profiles of electron–neutral collision cross sections are visualized for the six considered species, which demonstrate how these profiles can vary largely from one species to another and also show that this speed-dependence can be far from monotonic. (6) A mathematical description is given for how the speed-independent electron–neutral collision cross sections can be obtained from the speed-dependent profiles.

## 2. Research Method

This section aims to provide information about how the electron–neutral collision cross section values reported in the present study are obtained. After these values are obtained, they are reported in the form of plots showing their values versus the absolute electron temperature. Statistical analysis of the obtained data is then conducted to assess the spread of the collision cross sections obtained independently using four computing methods. The four methods to obtain the temperature-dependent, speed-independent electron–neutral collision cross sections for each of the six species covered herein are described next, in chronological order (the earliest first), based on the year of the published source corresponding to each method.

*2.1. Method 1 of 4, Simplified Frost Functions: $\bar{u}_e = 1.5\check{T}$*

The first method used herein to estimate the temperature-dependent electron–neutral collision cross sections is based on a published model [37] for computing the electric conductivity of plasma that may be encountered in magnetohydrodynamic (MHD) electric power generation. Such plasma is formed as combustion product gases seeded with an alkali metal vapor (either cesium, Cs, or potassium, K). The temperature range of the presented computational model was from 2000 K to 4000 K. In the published model for plasma electric conductivity, a list of expressions for the quantity ($\nu_{e-n}/N$) as functions of the monoenergetic (directly dependent on the instantaneous electron speed) electron energy ($u_e$) is given. These expressions were derived such that they approximately matched the corresponding experimental data available in the earlier literature, with one or two references identified for the source of data used to arrive at each expression. One expression was given for each of the 13 gaseous neutral species. The variable ($\nu_{e-n}$) represents the electron–neutral momentum transfer collision frequency for a particular species, and ($N$) represents the number density of that species in the plasma gas mixture. In the published electric conductivity model, the quantity ($\nu_{e-n}/N$) is equal to ($v_e Q$), which is the product of the electron speed and the electron–neutral momentum transfer cross section for the species of concern. Thus,

$$\nu_{e-n}(u_e) = v_e Q N \tag{8}$$

Solving for the momentum transfer collision cross section gives

$$Q(u_e) = \frac{\nu_{e-n}(u_e)}{v_e N} \tag{9}$$

or

$$Q(u_e) = \frac{\nu_{e-n}}{N}(u_e)\frac{1}{v_e} \tag{10}$$

This can be rewritten as

$$Q(u_e) = \frac{\psi(u_e)}{v_e} \tag{11}$$

where $\psi(u_e)$ is the electron-energy-dependent quantity ($\nu_{e-n}/N$), for which an expression was made in the published electric conductivity model.

In the present study, an approximation is made by utilizing Equation (11) with an average electron thermal speed of ($\overline{v}_e$) and an average electron kinetic energy ($\overline{u}_e$). The resulting electron–neutral momentum transfer collision cross section is designated as the simplified value ($\overline{Q}$), which is the approximated temperature-dependent alternative to the monoenergetic collision cross section. This simplified value of the collision cross section may be referred to as the "average", "mean", or "constant", with the meaning that it no longer depends on the distribution of the electron thermal speed and is not computed for a particular electron energy. Instead, it is interpreted as a constant approximate value of all the electron speeds and electron energies, which depends on the electron temperature as the sole independent variable. Thus,

$$\overline{Q}(T_e) = \frac{\psi(\overline{u}_e)}{\overline{v}_e} \tag{12}$$

In order to select proper values for the average electron thermal speed of ($\overline{v}_e$), and a proper average electron kinetic energy ($\overline{u}_e$), the Maxwell–Boltzmann probability distribution function for electron thermal speeds is employed. In this distribution for the thermal speed of electrons, the electron temperature ($T_e$) is the only adjustable parameter, as follows [38,39]:

$$f(v_e; T_e) = 4\pi v_e^2 \left(\frac{m_e}{2\pi k_B T_e}\right)^{1.5} \exp\left(-\frac{m_e v_e^2}{2 k_B T_e}\right) \tag{13}$$

The integral of $f(v_e; T_e)$ between two speed values ($v_{e1}$) and ($v_{e2}$), or $\int_{v_{e1}}^{v_{e2}} f(v_e; T_e) \, dv_e$, is the fraction of electrons with thermal speeds between ($v_{e1}$) and ($v_{e2}$), when the electrons have an absolute temperature of ($T_e$).

The mean thermal speed ($\overline{v}_e$) is related to the electron temperature ($T_e$) as

$$\overline{v}_e(T_e) = \sqrt{\frac{8 k_B}{\pi m_e}} \sqrt{T_e} \tag{14}$$

Inserting the values of the constants ($\pi$), ($k_B$), and ($m_e$) into the above equation gives the following relation:

$$\overline{v}_e \left[\frac{\text{m}}{\text{s}}\right] = 6,212.5114 \sqrt{T_e[\text{K}]} \tag{15}$$

with the electron kinetic energy defined as

$$u_e = \frac{1}{2} m_e v_e^2 \tag{16}$$

The mean kinetic energy of electrons based on the Maxwell–Boltzmann distribution is related to the electron temperature as

$$\overline{u}_e(T_e) = 1.5 k_B T_e \tag{17}$$

Inserting the value of the constant ($k_B$) into the above equation gives the following relation:

$$\overline{u}_e[\text{J}] = 2.0709735 \times 10^{-23} T_e[\text{K}] \tag{18}$$

If the mean electron energy is expressed in eV, the above equation becomes

$$\overline{u}_e[\text{eV}] = 1.5 \frac{k_B \left[\frac{\text{J}}{\text{K}}\right]}{e[\text{C}]} T_e[\text{K}] = 1.5 \hat{k}_B \left[\frac{\text{eV}}{\text{K}}\right] T_e[\text{K}] = 0.00012926000 \left[\frac{\text{eV}}{\text{K}}\right] T_e[\text{K}] \tag{19}$$

Using Equations (15) and (19) in Equation (12) gives

$$\overline{Q}(T_e)\left[\text{Å}^2\right] = 10^6 \frac{\psi(\overline{u}_e[\text{eV}])\left[10^{-8}\text{cm}^3/\text{s}\right]}{\overline{v}_e[\text{m/s}]} \tag{20}$$

The constant $10^6$ in the above equation was added to ensure that the units are consistent on both sides of the above equation. It is the result of multiplying a multiplier factor of $10^{20}$ to convert the area unit from square meters ($m^2$) to square angstroms ($\text{Å}^2$) by another multiplier factor of $10^{-14}$ to convert the volume unit from $10^{-8}$ cm$^3$ to m$^3$. Therefore, the above equation gives the average (simplified) collision cross sections, expressed in the unit of square angstroms ($\text{Å}^2 = 10^{-20} m^2$), which is appropriate for the scale of these small cross sections.

Using Equation (15) in Equation (20) gives

$$\overline{Q}(T_e)\left[\text{Å}^2\right] = \frac{10^6}{6,212.5114} \frac{\psi(\overline{u}_e[\text{eV}])\left[10^{-8}\text{cm}^3/\text{s}\right]}{\sqrt{T_e[\text{K}]}} \tag{21}$$

or

$$\overline{Q}(T_e)\left[\text{Å}^2\right] = 160.96550 \frac{\psi(\overline{u}_e[\text{eV}])\left[10^{-8}\text{cm}^3/\text{s}\right]}{\sqrt{T_e[\text{K}]}} \tag{22}$$

The above equation is the main expression needed to find the approximate electron–neutral collision cross section ($\overline{Q}$), in $\text{Å}^2$, for an arbitrary species defined by its function $\psi(\overline{u}_e[\text{eV}])$, in $10^{-8}$ cm$^3$/s, at a given electron absolute temperature $T_e$, in K. In the case of thermal equilibrium, electrons have the same temperature as the other plasma particles (molecules, atoms, and ions); thus, there is a single bulk (macroscopic) temperature to be used. This is a reasonable situation in seeded combustion gas plasma [40]. However, the analysis herein is not restricted to this situation, and thus, the electron temperature is not assumed to be equal to the bulk temperature of the plasma gas.

If the electron absolute temperature is expressed as an energy quantity in eV (and is designated herein as the symbol $\check{T}_e$), then the following transformation equation can be used to relate $\check{T}_e$ with $T_e$:

$$\check{T}_e[\text{eV}] = \frac{k_B[\text{J/K}]}{e[\text{C}]} T_e[\text{K}] \tag{23}$$

Using the above equation, Equation (17), namely, $\overline{u}_e(T_e) = 1.5 \, k_B T_e$, can be rewritten in a briefer form as

$$\overline{u}_e(T_e)[\text{eV}] = 1.5 \check{T}_e[\text{eV}] \tag{24}$$

Due to this relation, which is a key assumption used in implementing this method for computing the approximate electron–neutral collision cross sections, this particular method (method 1) is designated herein with the expression $\overline{u}_e = 1.5 \, \check{T}$ in the corresponding plots of the results.

The expressions for $\psi(u_e[\text{eV}])$, in $10^{-8}$ cm$^3$/s, for each of the six gaseous species considered herein are listed in Table 1. The species are ordered alphabetically, by chemical symbol.

**Table 1.** Published expressions for the quantity $\psi(u_e[\text{eV}])$ $\left[10^{-8}\text{cm}^3/\text{s}\right]$ in the first method.

| Index | Species | Mathematical Expression |
|---|---|---|
| 1 | CO | $\psi = 9.1 \, u_e$ |
| 2 | $CO_2$ | $\psi = \dfrac{1.7}{\sqrt{u_e}} + 2.1 \, u_e$ |
| 3 | $H_2$ | $\psi = 4.5 \sqrt{u_e} + 6.2 \, u_e$ |
| 4 | $H_2O$ | $\psi = \dfrac{10}{\sqrt{u_e}}$ |
| 5 | K | $\psi = 160$ |
| 6 | $O_2$ | $\psi = 2.75 \, u_e$ |

By replacing the monoenergetic ($u_e$) with the mean ($\bar{u}_e$), and then using Equations (19) and (22), the above expressions can be replaced with direct relations between the electron absolute temperature (in K) and the average electron–neutral collision cross section (in Å$^2$) for each species. These relations were derived and are provided in Table 2. These alternative expressions that use the electron temperature are much better than their counterparts that use the electron energy because the introduced alternative expressions allow for computing the average electron–neutral collision cross sections in a single step (without the need to first compute $\bar{u}_e$ and then $\psi$), and because they reveal the type of dependence of the average electron–neutral collision cross sections on the temperature. For example, after transforming the expressions of $\overline{Q}$ to be an explicit function of $T_e$, it becomes evident that none of the six considered species has its average electron–neutral collision cross section (as described according to method 1) depending exactly linearly on the absolute electron temperature. Instead, all the alternative expressions exhibit either a nonlinear correlation between the absolute electron temperature and the average electron–neutral collision cross section or a lack of any correlation at all (in the case of molecular oxygen, $O_2$).

**Table 2.** Derived expressions for the average electron–neutral collision cross section ($\overline{Q}$, in Å$^2$) in the first method as a direct function of the electron absolute temperature ($T_e$, in K).

| Index | Species | Mathematical Expression |
|---|---|---|
| 1 | CO | $\overline{Q} = 0.18933824\,\sqrt{T_e}$ |
| 2 | $CO_2$ | $\overline{Q} = 3.8431220 + \dfrac{24{,}068.535}{T_e}$ |
| 3 | $H_2$ | $\overline{Q} = 8.2352614 + 0.12899968\,\sqrt{T_e}$ |
| 4 | $H_2O$ | $\overline{Q} = \dfrac{141{,}579.62}{T_e}$ |
| 5 | K | $\overline{Q} = \dfrac{25{,}754.48}{\sqrt{T_e}}$ |
| 6 | $O_2$ | $\overline{Q} = 5.0326598$ |

*2.2. Method 2 of 4, Mori et al.'s Compiled Functions: Linear*

The second method used herein to estimate the temperature-dependent electron–neutral collision cross sections is based on a published study [41] regarding the theoretical calculation of fourteen thermodynamic and electric properties of combustion gases and their plasma for a range of temperatures from 0 K to 4000 K. In that study, expressions or constant values for the electron–neutral collision cross sections were listed for 23 species. All these published functions were either a zeroth-degree polynomial (a constant value) or a first-degree polynomial (a linear function). Thus, this method for computing the approximate electron–neutral collision cross sections is designated here by the word "Linear" in the corresponding plots of results, because its complexity does not go beyond a linear function in the electron absolute temperature. For the six covered gaseous species in the present study, the corresponding published temperature-dependent functions are listed in Table 3.

**Table 3.** Expressions for the average electron–neutral collision cross section ($\overline{Q}$, in Å$^2$) in the second method as a function of the electron absolute temperature ($T_e$, in K).

| Index | Species | Mathematical Expression |
|---|---|---|
| 1 | CO | $\overline{Q} = 8.6 + 0.0009\,(T_e - 2000)$ |
| 2 | $CO_2$ | $\overline{Q} = 30 - 0.01\,(T_e - 2000)$ |
| 3 | $H_2$ | $\overline{Q} = 11 + 0.001\,(T_e - 2000)$ |
| 4 | $H_2O$ | $\overline{Q} = 63 - 0.015\,(T_e - 2000)$ |
| 5 | K | $\overline{Q} = 250$ |
| 6 | $O_2$ | $\overline{Q} = 3.9$ |

*2.3. Method 3 of 4, Regression Fits for Raeder's Tabulated Results: Integration*

The third method to estimate the temperature-dependent electron–neutral collision cross sections is introduced herein by applying linear regression modeling to published numerical results [42] in the form of tabulated data. The raw results were for 9 gaseous species, covering electron temperature values from 2000 K to 4000 K, with a step of 100 K. The linear regression functions derived herein are based on the part of the raw data corresponding to the temperature range of interest only (between 2000 K and 3000 K) for the 6 covered species in the present study. The choice of linear regression is based on visually observing the trend of the plotted discrete data points (11 points per species), which was reasonably described with a straight line. The suitability of the linear fitting functions was later confirmed by computing the R-squared ($R^2$) value for each proposed linear regression function as a non-dimensional quantitative measure of the function's ability to reproduce the discrete data points. All the R-squared values were very close to their perfect upper limit of 1 [43–45], indicating excellent goodness of fit. Each published discrete value of the electron–neutral collision cross sections (at a given electron temperature) for each species is a result of an integration performed for the product of the monoenergetic version of the electron–neutral collision cross sections and a weight function ($g$), as follows:

$$\overline{Q}(T_e) = \int_0^\infty Q(v_e) g(x) dx \tag{25}$$

The variable ($x$) in the above equation is a non-dimensional version of the electron kinetic energy, after being normalized with the product of the Boltzmann constant and the absolute electron temperature ($k_B T_e$), which is called herein the "thermal energy" of the electrons. Thus, the non-dimensional variable ($x$) is a ratio of two energies (kinetic energy to thermal energy), or

$$x = \frac{0.5 m_e v_e^2}{k_B T_e} = \frac{u_e}{k_B T_e} \tag{26}$$

In the Maxwell–Boltzmann probability distribution of thermal speeds, the quantity ($2 k_B T_e / m_e$) is equal to the square of the most-probable speed ($v_p$), at which this probability distribution has its peak. Thus,

$$v_p^2 = \frac{2 k_B T_e}{m_e} \tag{27}$$

Or

$$k_B T_e = 0.5 m_e v_p^2 \tag{28}$$

Using Equation (28) in Equation (26), the integration variable ($x$) can be viewed as the square of the ratio of the electron speed to its most-probable speed, or

$$x = \frac{0.5 m_e v_e^2}{0.5 m_e v_p^2} = \left(\frac{v_e}{v_p}\right)^2 \tag{29}$$

Alternatively, for a known value of ($x$), the corresponding value of $v_e$ can be calculated, provided that the electron temperature is also specified, according to

$$v_e = v_p(T_e) \sqrt{x} = \sqrt{\frac{2 k_B T_e}{m_e}} \sqrt{x} \tag{30}$$

Thus, for a given electron absolute temperature ($T_e$) that is treated as a fixed parameter when performing the integration in Equation (25), that integration can easily be viewed as being performed with respect to the non-dimensional variable ($x$) instead of the dimensional monoenergetic electron speed ($v_e$), as

$$\overline{Q}(T_e) = \int_0^\infty Q(x; T_e) g(x) dx \tag{31}$$

This means that the electron temperature implicitly affects the integral through $v_p$ or $x$. The weight function $g(x)$ has the following form:

$$g(x) = 0.5 x^2 e^{-x} \tag{32}$$

The value of $g$ is 0 at $x = 0$. It has a single peak, with a maximum value of 0.27067 at $x = 2$. The value of $g$ approaches 0 at the limit of $x \to \infty$. This means that practically, the integration in Equation (31) can be performed over a finite range of $x$ up to approximately $x = 10$ (where $g$ diminishes to 0.00227).

Likewise, the integration variable can be changed in another way by utilizing the linear relation between the monoenergetic kinetic energy of the electron ($u_e = 0.5\ m_e v_e^2$) and the non-dimensional variable ($x$) at a given electron temperature, as

$$u_e[\text{J}] = x k_B T_e[\text{J}] \text{ (if } u_e \text{ is expressed in J)} \tag{33}$$

or

$$u_e[\text{eV}] = x \hat{k}_B T_e[\text{eV}] \text{ (if } u_e \text{ is expressed in eV)} \tag{34}$$

Thus,

$$x = \frac{u_e[\text{J}]}{k_B T_e[\text{J}]} = \frac{u_e[\text{eV}]}{\hat{k}_B T_e[\text{eV}]} = \frac{u_e[\text{eV}]}{\check{T}_e[\text{eV}]} \tag{35}$$

Therefore, the integration for computing the average (speed-independent) electron–neutral cross section $\overline{Q}(T_e)$ in the current method (third method) can be viewed as being performed with respect to the monoenergetic electron energy. Specifically, using Equation (35) in Equation (31) gives

$$\overline{Q}(T_e) = \int_0^\infty Q\left(\frac{u_e[\text{J}]}{k_B T_e}; T_e\right) g\left(\frac{u_e[\text{J}]}{k_B T_e}\right) d\left(\frac{u_e[\text{J}]}{k_B T_e}\right) = \int_0^\infty Q\left(\frac{u_e[\text{eV}]}{\hat{k}_B T_e}; T_e\right) g\left(\frac{u_e[\text{eV}]}{\hat{k}_B T_e}\right) d\left(\frac{u_e[\text{eV}]}{\hat{k}_B T_e}\right) \tag{36}$$

Or, with either ($k_B T_e$) or ($\hat{k}_B T_e$) treated as an invariant during the integration,

$$\overline{Q}(T_e) = \frac{1}{k_B T_e} \int_0^\infty Q(u_e[\text{J}]; T_e) g\left(\frac{u_e[\text{J}]}{k_B T_e}\right) du_e[\text{J}] = \frac{1}{\hat{k}_B T_e} \int_0^\infty Q(u_e[\text{eV}]; T_e) g\left(\frac{u_e[\text{eV}]}{\hat{k}_B T_e}\right) du_e[\text{eV}] \tag{37}$$

In the discussed method 3 herein, the monoenergetic values of the electron–neutral cross sections ($Q$) in the published source study of this method were obtained for each species according to published information in one or more external reference sources, or according to the extrapolation of the published information, such that a range of monoenergetic electron kinetic energy ($u_e$) values from 0.0005 eV to 6 eV is covered for each species.

Using Equation (35), namely, $x = u_e[\text{eV}]/\hat{k}_B T_e[\text{eV}]$, the corresponding ranges of ($x$) at the lower electron temperature limit (2000 K) and at the upper electron temperature limit (3000 K) considered in the present study can be obtained. Furthermore, Equation (30) allows for determining the corresponding lower and upper limits of the monoenergetic electron thermal speed in the integrations performed. Alternatively, a modified version of Equation (16) can be used to find the monoenergetic electron velocity from a known monoenergetic electron kinetic energy, as

$$v_e\left[\frac{\text{m}}{\text{s}}\right] = \sqrt{\frac{2 u_e[\text{J}]}{m_e[\text{kg}]}} = \sqrt{\frac{2 u_e[\text{eV}] e[\text{C}]}{m_e[\text{kg}]}} \tag{38}$$

The results are provided in Table 4.

**Table 4.** Mapping the limits of the monoenergetic electron kinetic energy ($u_e$) to the monoenergetic electron thermal speed ($v_e$); to the non-dimensional energy ratio ($x$) at $T_e$ = 2000 K; and to the non-dimensional energy ratio ($x$) at $T_e$ = 3000 K for the integrations implied in method 3.

| Limit Value | $T_e$ = 2000 K | $T_e$ = 3000 K |
|---|---|---|
| Smallest $u_e$ (eV) | 0.0005 | |
| Largest $u_e$ (eV) | 6.0000 | |
| Smallest $v_e$ (m/s) at smallest $u_e$ | 13,262 | |
| Largest $v_e$ (m/s) at largest $u_e$ | 1,452,785 | |
| Smallest $x$ (-) at smallest $u_e$ | 0.00258 | 0.00172 |
| Largest $x$ (-) at largest $u_e$ | 34.81355 | 23.20904 |

Table 5 demonstrates the stretching (or spread) of the $g$ weight function upon elevating the electron temperature if this weight function is expressed in terms of the monoenergetic electron kinetic energy or the monoenergetic electron thermal speed. This is achieved by comparing the location of the peak of $g$ (which is 0.270671 at $x$ = 2) in terms of the monoenergetic electron kinetic energy ($u_e$) and in terms of the corresponding monoenergetic electron thermal speed ($v_e$), at absolute electron temperatures ($T_e$) of 2000 K and 3000 K. The peak $g$ is shifted to a new (larger) location of $u_e$ according to the ratio of the absolute electron temperatures; thus, the $u_e$ location increased by a factor of 3000/2000 = 1.5. The peak $g$ is shifted to a new (larger) location of $v_e$ according to the ratio of the square root of the absolute electron temperatures; thus, the $v_e$ location increases by a factor of $\sqrt{3000/2000} = \sqrt{1.5} = 1.2247$. It should be noted that the peak of $g$ itself is the same (remains 0.270671) during this horizontal-axis-only stretching.

**Table 5.** Shift of the peak $g$ weight function when expressed as dependent on the monoenergetic electron kinetic energy ($u_e$) or on the monoenergetic electron thermal speed ($v_e$), as the electron absolute temperature increases from 2000 K to 3000 K.

| Value at Peak $g$ (=0.270671 [-]) | $T_e$ = 2000 K | $T_e$ = 3000 K |
|---|---|---|
| $x$ (-) at peak $g$ | 2.0000 | |
| $u_e$ (eV) at peak $g$ | 0.34469 | 0.51704 |
| $v_e$ (m/s) at peak $g$ | 348,211 | 426,469 |

The obtained linear regression functions based on the published data points of the integration-based electron–neutral collision cross section for each of the six covered species herein are listed in Table 6. The R-squared ($R^2$) value is also listed for each regression model.

**Table 6.** Proposed linear regression models for the average electron–neutral collision cross section ($\overline{Q}$, in Å$^2$) in the third method as a function of the electron absolute temperature ($T_e$, in K), based on fitting 11 published data points for each species. The obtained non-dimensional goodness-of-fit R-squared ($R^2$) value for each proposed linear regression model is also provided.

| Index | Species | Linear Regression Equation | R-Squared |
|---|---|---|---|
| 1 | CO | $\overline{Q} = 7.840000 + 0.003098\ T_e$ | 0.99916 |
| 2 | $CO_2$ | $\overline{Q} = 24.431818 - 0.005240\ T_e$ | 0.98317 |
| 3 | $H_2$ | $\overline{Q} = 12.512727 + 0.000955\ T_e$ | 0.99128 |
| 4 | $H_2O$ | $\overline{Q} = 107.955455 - 0.023073\ T_e$ | 0.98272 |
| 5 | K | $\overline{Q} = 362.090909 - 0.068000\ T_e$ | 0.99326 |
| 6 | $O_2$ | $\overline{Q} = 4.010000 + 0.000767\ T_e$ | 0.99787 |

*2.4. Method 4 of 4, Bedick et al.'s Recommended Correlations: Quadratic*

The fourth and last method used herein to estimate the temperature-dependent electron–neutral collision cross sections is based on a recommended set of published polynomials [46], with the single independent variable being the electron absolute temperature for 15 gaseous species. These published polynomials were made based on data analyzed in the literature, with one or more reference sources given for each species. These polynomial functions were up to a third degree for a range of temperatures from 1500 K to 3500 K. For the six particular species covered herein, quadratic polynomials were proposed for four of them ($CO_2$, $H_2$, $H_2O$, and K), while linear polynomials were proposed for the other two species (CO and $O_2$). Therefore, none of the six species required a cubic polynomial. Thus, this method for computing the approximate electron–neutral collision cross sections is designated herein by the word "Quadratic" in the corresponding plots of the results. For the six covered gaseous species in the present study, the published temperature-dependent correlations are listed in Table 7.

**Table 7.** Recommended correlations for the average electron–neutral collision cross section ($\overline{Q}$, in Å$^2$) in the fourth method as a function of the electron absolute temperature ($T_e$, in K).

| Index | Species | Correlation |
|---|---|---|
| 1 | CO | $\overline{Q} = 5.523 + 0.001913\, T_e$ |
| 2 | $CO_2$ | $\overline{Q} = 79.82 - 0.02871\, T_e + 3.148 \times 10^{-6}\, T_e^2$ |
| 3 | $H_2$ | $\overline{Q} = 9.410 + 0.001287\, T_e - 6.578 \times 10^{-8}\, T_e^2$ |
| 4 | $H_2O$ | $\overline{Q} = 406.6 - 0.1842\, T_e + 2.533 \times 10^{-5}\, T_e^2$ |
| 5 | K | $\overline{Q} = 958.2 - 0.3434\, T_e + 4.291 \times 10^{-5}\, T_e^2$ |
| 6 | $O_2$ | $\overline{Q} = 2.569 + 6.473 \times 10^{-4}\, T_e$ |

## 3. Results

*3.1. Demonstration of Monoenergetic Collision Cross Sections*

The results section starts here with demonstrations of the published discrete-points monoenergetic collision cross sections corresponding to method 3, along with the weight function for each of the six species covered herein. Although this part is not the main outcome of the present study, it is still considered very helpful in explaining the influence of the electron temperature and the electrons' monoenergetic thermal speed (thus, the electrons' monoenergetic kinetic energy) on the predicted simplified speed-independent electron–neutral collision cross sections.

Figure 1 shows a graphical representation of the tabulated data of the monoenergetic (speed-dependent) electron–neutral collision cross sections ($Q$) for carbon monoxide (CO). The figure combines this graphical representation of the collision cross sections with another graphical representation for the corresponding weight function ($g$), which was computed here after the value of the non-dimensional energy ratio ($x$) was deduced for each of the published values of the electron kinetic energy ($u_e$). For each of the six gaseous species, the monoenergetic collision cross sections are strictly dependent on the electron monoenergetic thermal speed ($v_e$) or, equivalently, on the related electron monoenergetic kinetic energy ($u_e$). However, these monoenergetic collision cross sections can also be viewed as a function of the non-dimensional energy ratio ($x$) but for a given electron absolute temperature ($T_e$). Thus, three scales are used in the horizontal axis in the figure, such that each of these three variables can be utilized as the independent variable for $Q$. In the first (top) plot of the figure, $T_e$ = 2000 K. This parameter is necessary to determine the value of $x$ at each value of $u_e$ or $v_e$. In the second (bottom) plot of the figure, the $x$ values are inferred from the $u_e$ or $v_e$ values with a higher parameter value of $T_e$ = 3000 K. The change in $T_e$ does not impact the profile of $Q(u_e)$ or $Q(v_e)$ because these profiles are temperature-invariant. Instead, the increase in $T_e$ only impacts the profile of $g(u_e; T_e)$ or $g(v_e; T_e)$ by stretching the profile of $g$ horizontally along the horizontal $u_e$ axis or the horizontal $v_e$ axis, respectively.

Despite this, the profile of $g(x)$ is the same in both plots, because the relation between $g$ and $x$ is temperature-invariant, as previously shown in Equation (32). The figure illustrates the role of $T_e$ in broadening the profile of $g(u_e; T_e)$ or $g(v_e; T_e)$ relative to the static profile of $Q(u_e)$ or $Q(v_e)$, leading to different integration results due to the deformation in the weight function, not due to a change in the monoenergetic collision cross section function. The original data were published for a range of monoenergetic electron kinetic energy ($u_e$) values from 0.0005 eV to 6.000 eV. However, the ranges of the horizontal $u_e$ axis in the figure plots extend slightly beyond this range in order to have a convenient and appealing logarithmic-scale range in terms of the interrelated axis of the electron thermal speed ($v_e$), but the ranges of the data plotted are exactly limited to those in the published data.

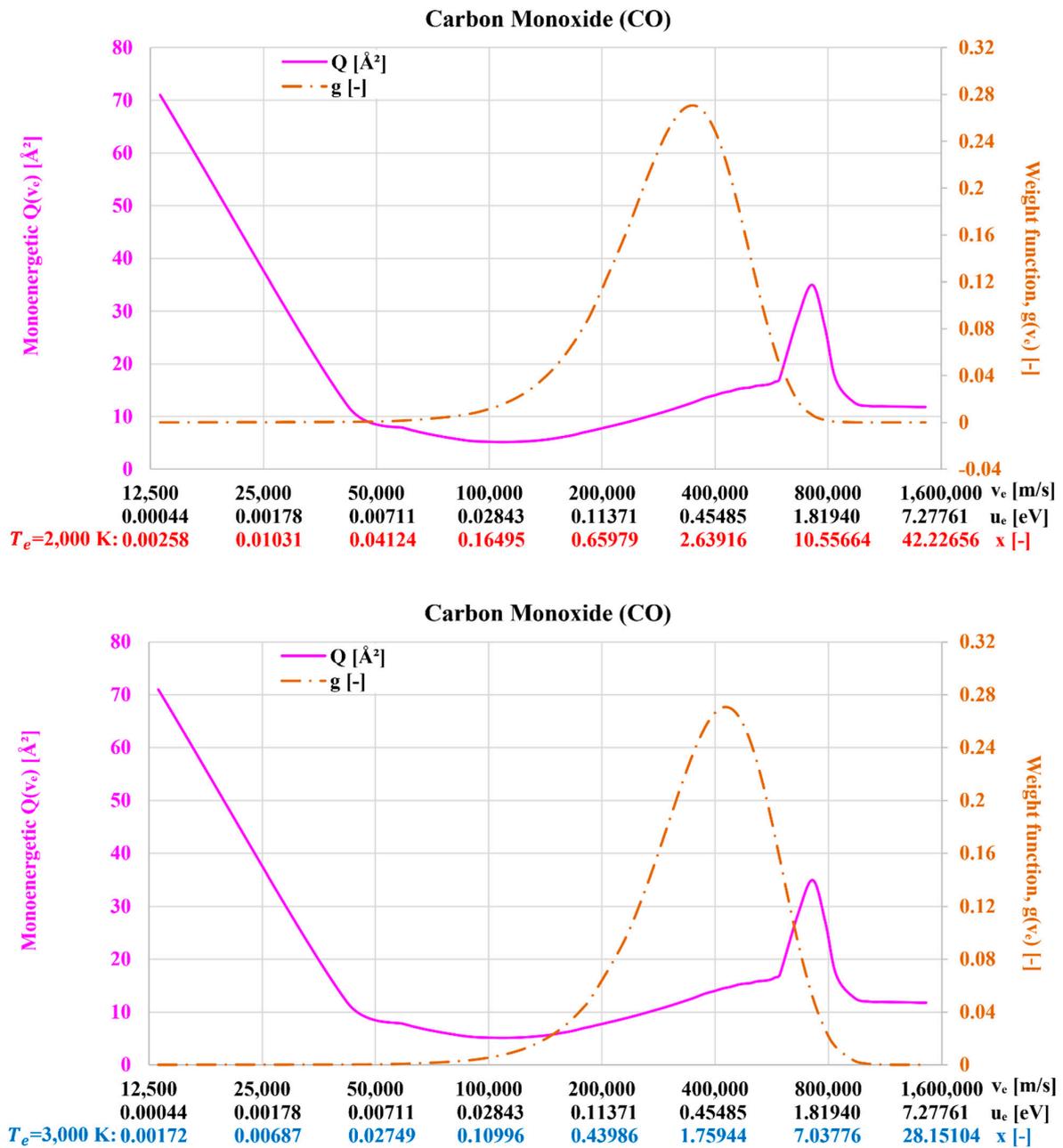

**Figure 1.** Profiles of the monoenergetic electron–neutral collision cross sections and the corresponding weight function for carbon monoxide (CO). The $(x - u_e)$ mapping and the $(x - v_e)$ mapping correspond to an absolute electron temperature of 2000 K in the top plot and to an absolute electron temperature of 3000 K in the bottom plot.

Figure 2 shows a similar graphical representation of the tabulated data of the monoenergetic electron-neutral collision cross-sections ($Q$) for carbon dioxide ($CO_2$), combined with the computed corresponding weight function ($g$) at $T_e$ = 3000 K (as an example). Figures 3–6 have similar graphical representations for molecular hydrogen ($H_2$), water vapor ($H_2O$), potassium vapor (K), and molecular oxygen ($O_2$); respectively. These figures clarify how the profiles of the monoenergetic collision cross-sections can vary significantly from one gaseous species to another, which also explains how the same weight function can give very different behaviors for different species, in terms of the dependence of the simplified (weighed average, speed-independent) collision cross-sections ($\overline{Q}$) on the absolute electron temperature ($T_e$).

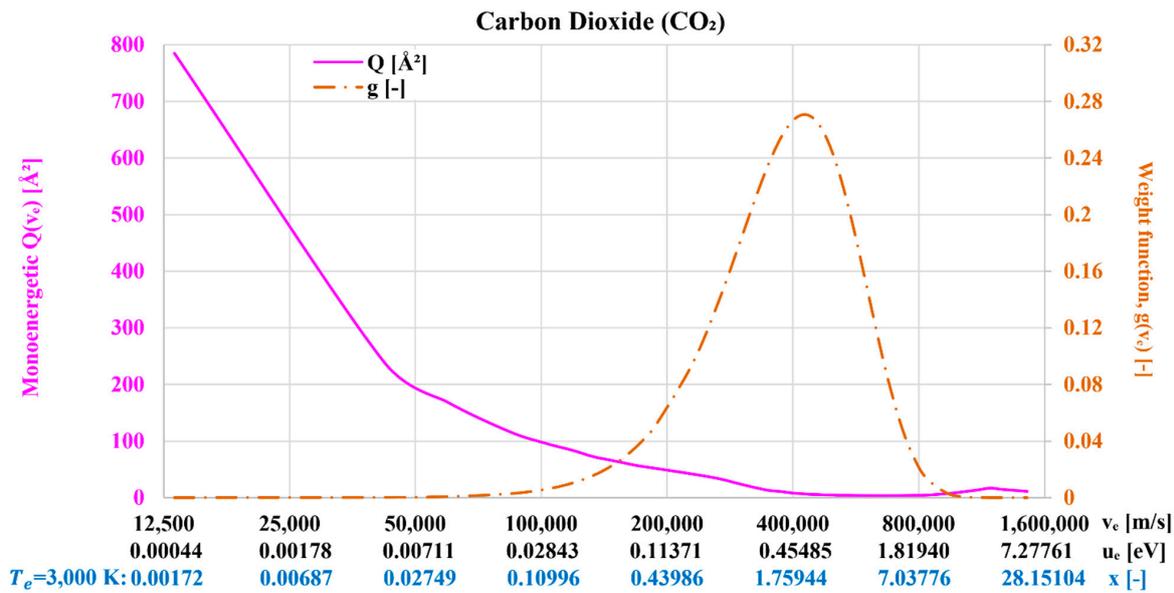

**Figure 2.** Profiles of the monoenergetic electron–neutral collision cross section and the corresponding weight function for carbon dioxide ($CO_2$). The ($x - u_e$) mapping and the ($x - v_e$) mapping correspond to an absolute electron temperature of 3000 K.

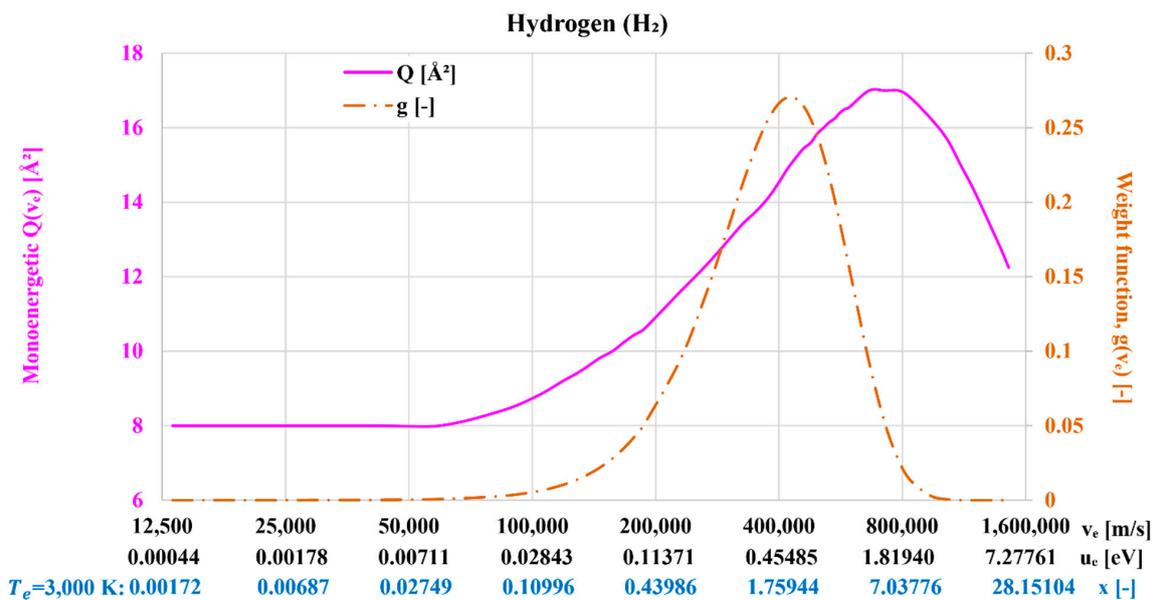

**Figure 3.** Profiles of the monoenergetic electron–neutral collision cross section and the corresponding weight function for molecular hydrogen ($H_2$). The ($x - u_e$) mapping and the ($x - v_e$) mapping correspond to an absolute electron temperature of 3000 K.

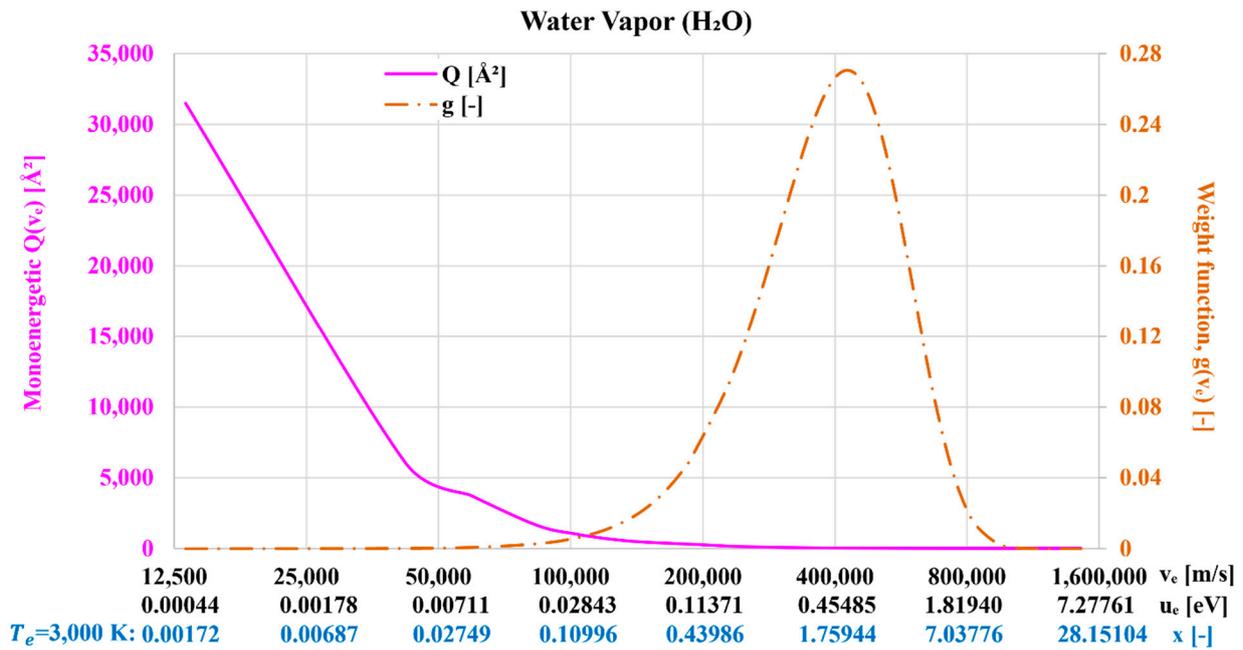

**Figure 4.** Profiles of the monoenergetic electron–neutral collision cross section and the corresponding weight function for water vapor ($H_2O$). The ($x - u_e$) mapping and the ($x - v_e$) mapping correspond to an absolute electron temperature of 3000 K.

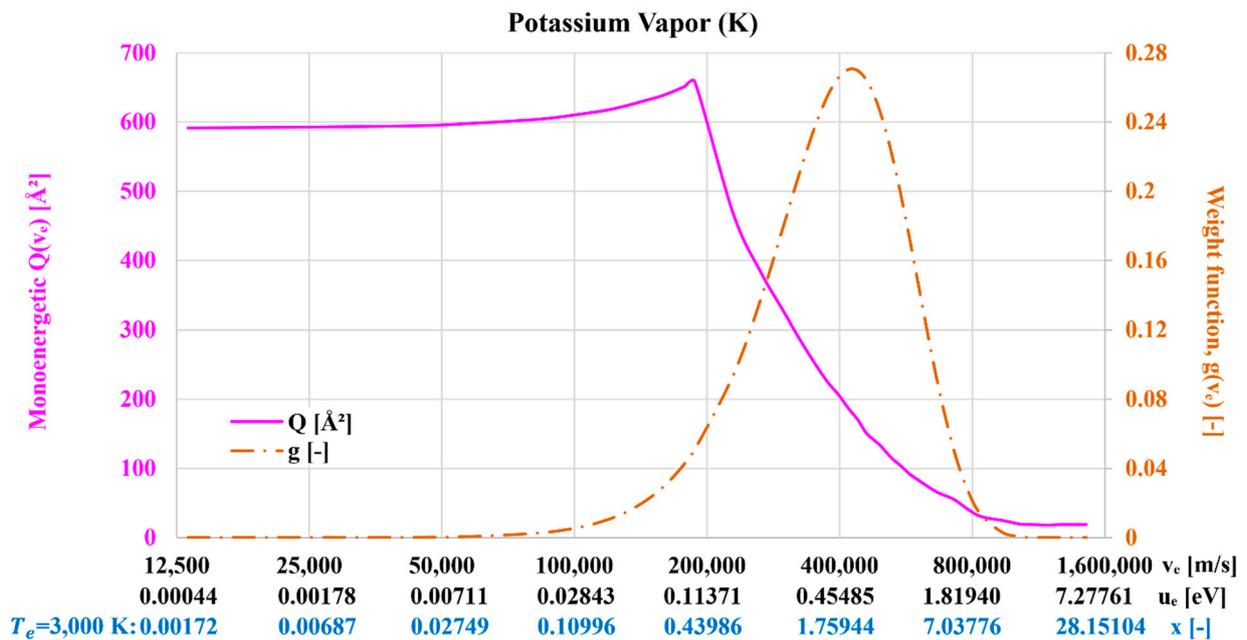

**Figure 5.** Profiles of the monoenergetic electron–neutral collision cross section and the corresponding weight function for potassium vapor (K). The ($x - u_e$) mapping and the ($x - v_e$) mapping correspond to an absolute electron temperature of 3000 K.

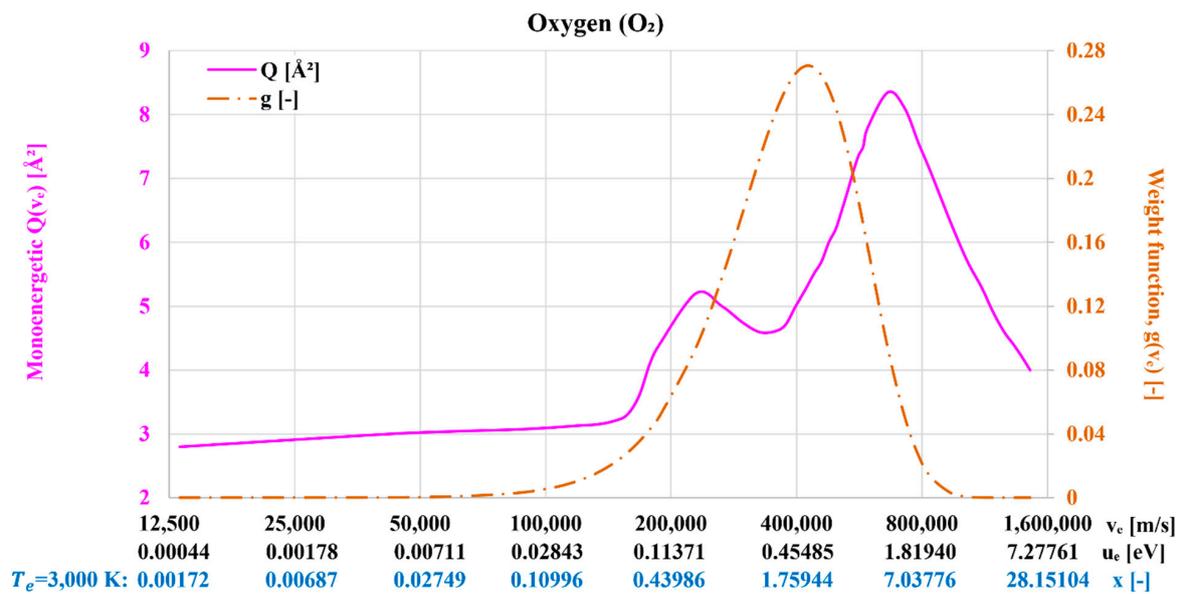

**Figure 6.** Profiles of the monoenergetic electron–neutral collision cross section and the corresponding weight function for molecular oxygen ($O_2$). The ($x - u_e$) mapping and the ($x - v_e$) mapping correspond to an absolute electron temperature of 3000 K.

*3.2. Temperature Profiles of Average Collision Cross Sections*

In this part, the dependence of the average electron–neutral collision cross section ($\overline{Q}$) on the electron temperature ($T_e$) for each of the six covered gaseous species is visualized for the range of $T_e$ from 2000 K to 3000 K. For each species, a figure is provided where four curves representing the obtained $\overline{Q}$ profiles using the four methods discussed earlier are displayed. This not only helps contrast the predictions using the different methods but also helps reveal the general trend of $\overline{Q}$ as $T_e$ increases from the lower temperature limit of 2000 K (typical of air–fuel combustion) to the upper temperature limit of 3000 K (typical of oxy–fuel combustion). For the third method (integration-based weighted average), the published discrete values are plotted as scattered markers, while a dotted straight line is added near them that represents the regression model derived here for these data points.

Figure 7 is used to contrast the four predicted curves of the average electron–neutral collision cross sections for carbon monoxide (CO). All four methods agree about the increase in $\overline{Q}_{CO}$ as $T_e$ increases. Based on the arithmetic mean profile (not shown; it passes through the points representing the arithmetic mean, also called the simple average, of the four $\overline{Q}$ values using the four methods at each temperature), the mean rate of change in $\overline{Q}_{CO}$ is 0.0019515 Å$^2$/K. This is the average change in $\overline{Q}_{CO}$ as $T_e$ increases by 1 K. Methods 1 ($\overline{u}_e = 1.5\check{T}_e$), 2 (linear), and 4 (quadratic) have similar values of $\overline{Q}_{CO}$, which together deviate remarkably from the results of method 3 (integration). Method 3 gives a larger $\overline{Q}_{CO}$ for the entire examined range of $T_e$. Although method 1 involves the use of a nonlinear function for $\overline{Q}_{CO}(T_e)$, the nonlinearity in the temperature range of interest here is too weak to be noticeable. Method 4 proposes a linear (not a quadratic) modeling function for $\overline{Q}_{CO}(T_e)$; thus, the displayed perfectly straight line obtained via this method is expected.

Figure 8 is used to contrast the four predicted curves of the average electron–neutral collision cross sections for carbon dioxide ($CO_2$). All four methods agree about the decrease in $\overline{Q}_{CO2}$ as $T_e$ increases (opposite behavior to $\overline{Q}_{CO}$). Based on the arithmetic mean profile of the four individual profiles, the mean rate of change in $\overline{Q}_{CO2}$ is −0.0080654 Å$^2$/K. The predictions of methods 1 ($\overline{u}_e = 1.5\check{T}_e$) and 3 (integration) have similar values of $\overline{Q}_{CO2}$, which together are noticeably smaller than those predicted using methods 2 (linear) and 4 (quadratic). The ratio of the largest $\overline{Q}_{CO2}$, obtained using method 4, to the smallest $\overline{Q}_{CO2}$, obtained using method 3, is 2.4385 at 2000 K, and 2.4280 at 3000 K. The values of $\overline{Q}_{CO2}$ are nearly twice those of $\overline{Q}_{CO}$; thus, carbon dioxide has a stronger interaction with electrons than carbon monoxide, causing less electron mobility and more suppression of the plasma electric conductivity. Despite the nonlinear expression for $\overline{Q}_{CO2}(T_e)$ in methods 1 and 4, the actual predictions in the temperature range of interest here resemble a linear decline when the electron temperature increases, and thus, the nonlinearity is not strongly manifested.

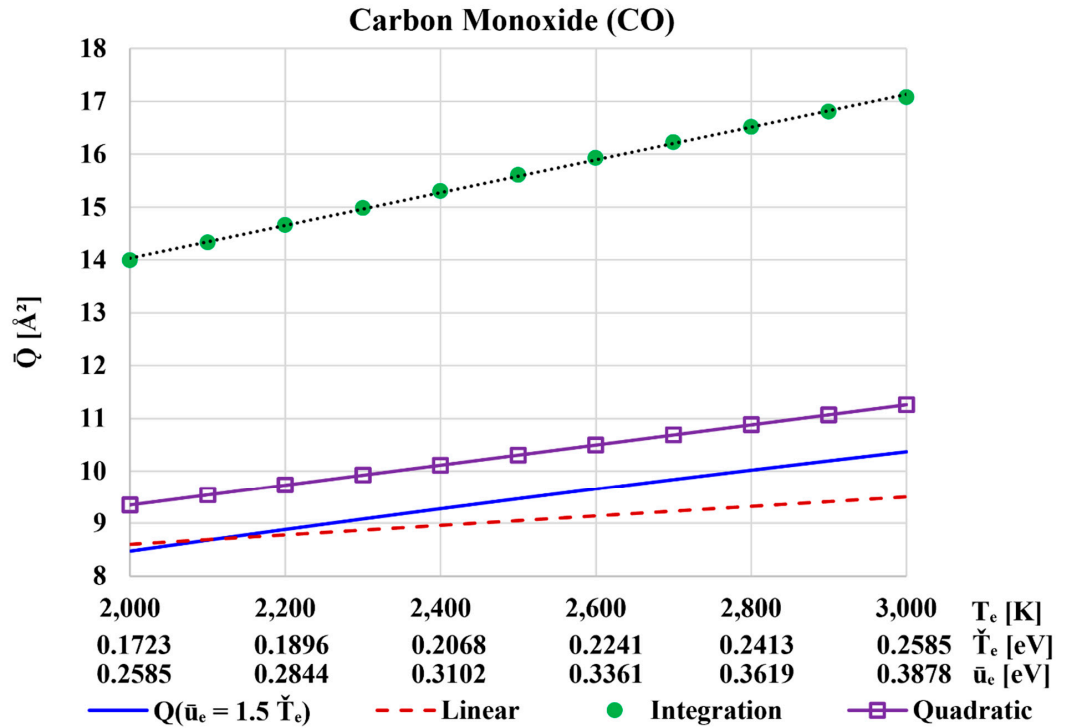

**Figure 7.** Temperature dependence of the average electron–neutral collision cross sections of carbon monoxide (CO), as obtained using four different methods.

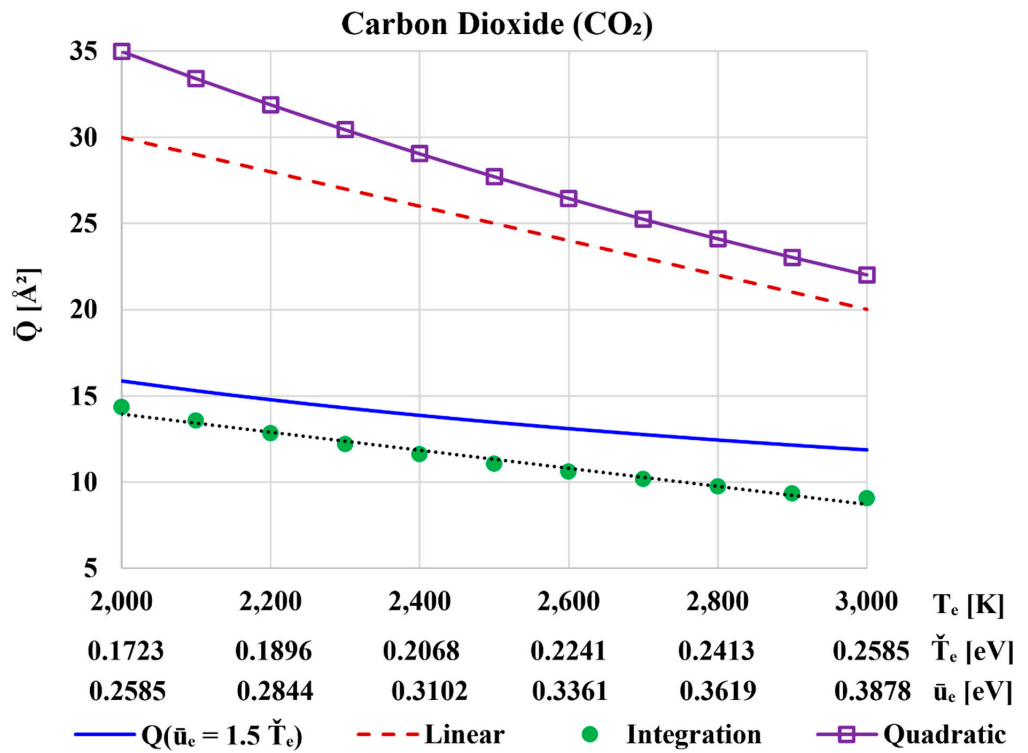

**Figure 8.** Temperature dependence of the average electron–neutral collision cross sections of carbon dioxide ($CO_2$), as obtained using four different methods.

Figure 9 is used to contrast the four predicted curves of the average electron–neutral collision cross sections for molecular hydrogen ($H_2$). All four methods agree about the increase in $\overline{Q}_{H2}$ as $T_e$ increases (the opposite behavior to $\overline{Q}_{CO2}$). Based on the arithmetic mean profile of the four individual profiles, the mean rate of change in $\overline{Q}_{H2}$ is 0.0010537 Å$^2$/K. The predictions of methods 1 ($\bar{u}_e = 1.5\check{T}_e$) and 3 (integration) have similar values of $\overline{Q}_{H2}$, which together are noticeably larger than those predicted using methods 2 (linear) and 4 (quadratic). The distinction of the predictions into these two groups is the same situation observed for $CO_2$, but in the case of $CO_2$, the predictions of methods 1 and 3 are the lower-values group (not the higher-values group, as in the case of $H_2$). Also, the gap between the two groups of predictions in the case of $H_2$ is not very large. For example, the ratio of the largest $\overline{Q}_{H2}$, obtained using method 3, to the smallest $\overline{Q}_{H2}$, obtained using method 2, is 1.3064 at 2000 K, and 1.2775 at 3000 K. The values of $\overline{Q}_{H2}$ are comparable to those of $\overline{Q}_{CO}$. As in the case of $CO_2$, the nonlinearity of $\overline{Q}_{H2}(T_e)$ in methods 1 and 4 is not pronounced in the temperature range of interest here. This is similar to the case of $CO_2$, except that $\overline{Q}_{CO2}(T_e)$ increases, not decreases, with ($T_e$).

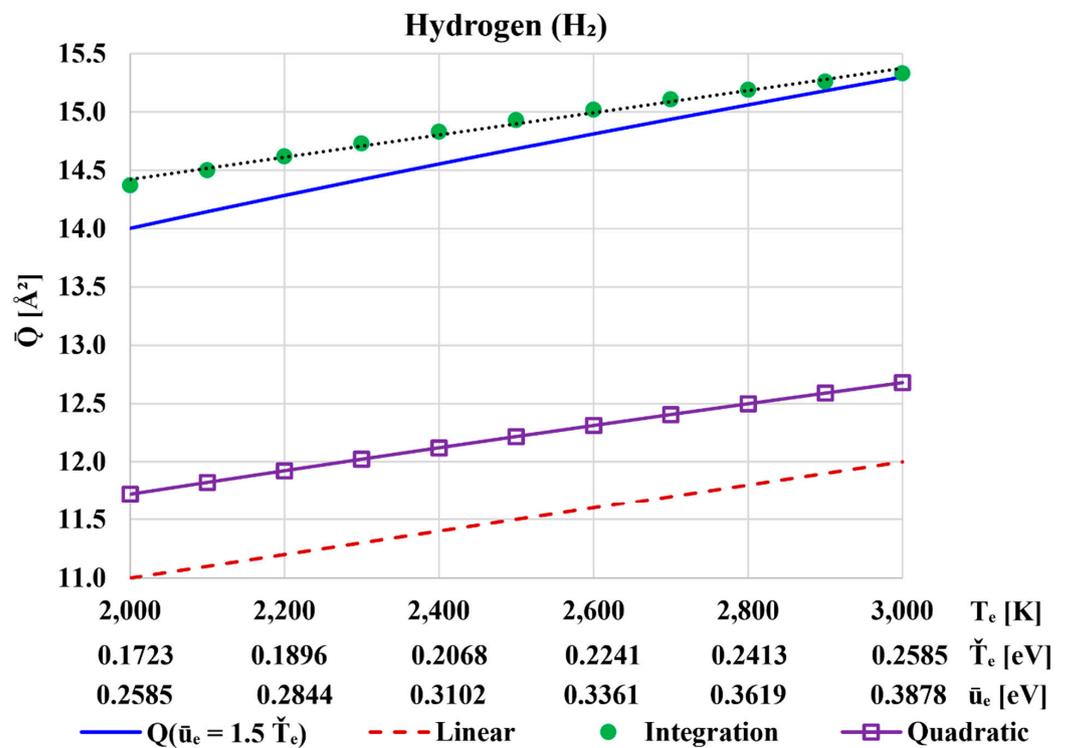

**Figure 9.** Temperature dependence for the average electron–neutral collision cross sections of molecular hydrogen ($H_2$), as obtained using four different methods.

Figure 10 is used to contrast the four predicted curves of the average electron–neutral collision cross sections for water vapor ($H_2O$). All four methods agree about the decrease in $\overline{Q}_{H2O}$ as $T_e$ increases (the opposite behavior to $\overline{Q}_{H2}$). Based on the arithmetic mean profile of the four individual profiles, the mean rate of change in $\overline{Q}_{H2O}$ is −0.0298717 Å$^2$/K. The predictions of methods 1 ($\bar{u}_e = 1.5\check{T}_e$), 2 (linear), and 3 (integration) have similar values of $\overline{Q}_{H2O}$, which together are noticeably smaller than those predicted using method 4 (quadratic). The observation that one method of the four behaves very differently than the other three methods in terms of the values of $\overline{Q}$ (but not in terms of its trend) was also found for CO, but in that case of CO, the outlier was method 3, not method 4. The values of $\overline{Q}_{H2O}$ are one order of magnitude larger than those of $\overline{Q}_{CO}$ and $\overline{Q}_{H2}$. This is an important feature of $H_2O$, indicating that combustion plasma produced from burning hydrogen-rich fuels is expected to have a smaller plasma electric conductivity compared with combustion plasma produced from burning carbon-rich fuels, if all other conditions are the same and if the additional effect of electron–ion and electron–electron scattering (Coulomb scattering) is equal. For example, when comparing the hot gaseous combustion products of hydrogen and the hot gaseous combustion products of carbon-rich coal, the former gases are expected to have a lower electric conductivity than the latter gases, assuming the same seeding amounts and same temperatures, as well as assuming the ions and electrons' effects on electron mobility are equal in both cases. In the case of $H_2O$, the nonlinearity involved in methods 1 and 4 is at its strongest level among all six species covered here. The magnitude of the negative slope of $\overline{Q}_{H2O}(T_e)$, or $d\overline{Q}_{H2O}/dT_e$, becomes smaller as $(T_e)$ increases, and the curvature of the respective profile curves can be visually noticed.

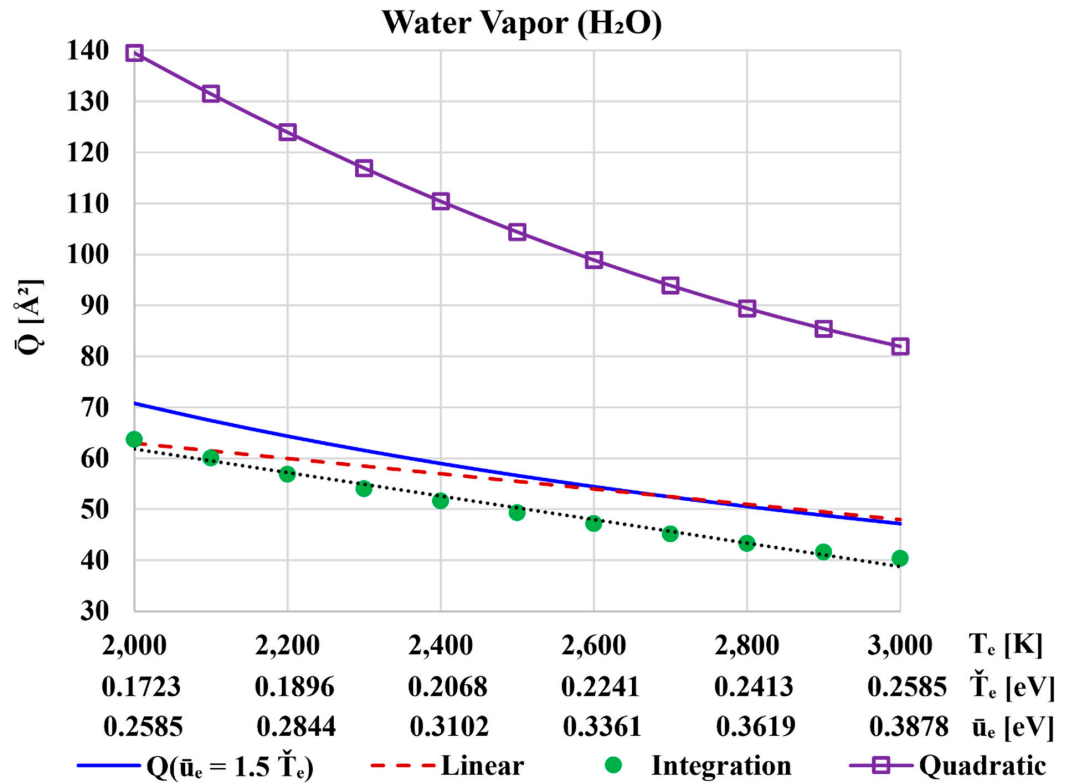

**Figure 10.** Temperature dependence for the average electron–neutral collision cross sections of water vapor ($H_2O$), as obtained using four different methods.

The slope (the first derivative) of $\overline{Q}_{H2O}(T_e)$ with respect to $T_e$ in method 1 ($\overline{u}_e = 1.5\check{T}_e$) is

$$\left.\frac{d\overline{Q}_{H2O}}{dT}\right|_{\overline{u}_e=1.5\check{T}_e} \left[\frac{\text{Å}^2}{\text{K}}\right] = -\frac{141,579.62}{T_e[\text{K}]^2} \tag{39}$$

For $T_e$ = 2000 K, the above slope is $-0.035395$ Å$^2$/K, which becomes $-0.015731$ Å$^2$/K at 3000 K. Thus, the magnitude of the slope drops at 3000 K to 44.44% of its value at 2000 K. The slope of $\overline{Q}_{H2O}(T_e)$ with respect to $T_e$ in method 4 (quadratic) is

$$\left.\frac{d\overline{Q}_{H2O}}{dT}\right|_{\text{Quadratic}} \left[\frac{\text{Å}^2}{\text{K}}\right] = -0.1842 + 5.066 \times 10^{-5} T_e[\text{K}] \tag{40}$$

For $T_e$ = 2000 K, the above slope is $-0.082880$ Å$^2$/K, which becomes $-0.032220$ Å$^2$/K at 3000 K. Thus, the magnitude of the slope drops at 3000 K to 38.88% of its value at 2000 K.

Figure 11 is used to contrast the four predicted curves of the average electron–neutral collision cross sections for potassium vapor (K). Three of the four methods agree about the decrease in $\overline{Q}_K$ as $T_e$ increases (similar behavior to $\overline{Q}_{H2O}$). The outlier is method 2 (linear), which estimates that $\overline{Q}_K$ is simply a constant value of 250 Å$^2$, independent of the electron temperature. For the three methods predicting a decrease in $\overline{Q}_K$ as $T_e$ increases, there is a large gap between each pair of them, with method 1 ($\overline{u}_e = 1.5\check{T}_e$) giving the largest values of $\overline{Q}_K$, and method 3 (integration) gives the smallest values. At $T_e$ = 2000 K, the computed $\overline{Q}_K$ using method 1 is 575.888 Å$^2$, while method 3 gives 229 Å$^2$; thus, the ratio is 2.5148:1. At $T_e$ = 3000 K, the computed $\overline{Q}_K$ using method 1 is 470.210 Å$^2$, while method 3 gives 161 Å$^2$; thus, the ratio is 2.9206:1. Based on the arithmetic mean profile of the four individual profiles (including the zero-degree polynomial of method 2), the mean rate of change in $\overline{Q}_K$ is $-0.0756318$ Å$^2$/K. The values of $\overline{Q}_K$ are the largest among the six species covered in the present study, being much larger than the values of $\overline{Q}_{H2O}$, which are themselves much larger than the values of $\overline{Q}_{CO}$, $\overline{Q}_{CO2}$, and $\overline{Q}_{H2}$. As in the cases of $CO_2$ and $H_2$, the nonlinearity of $\overline{Q}_K(T_e)$ in methods 1 and 4 is small.

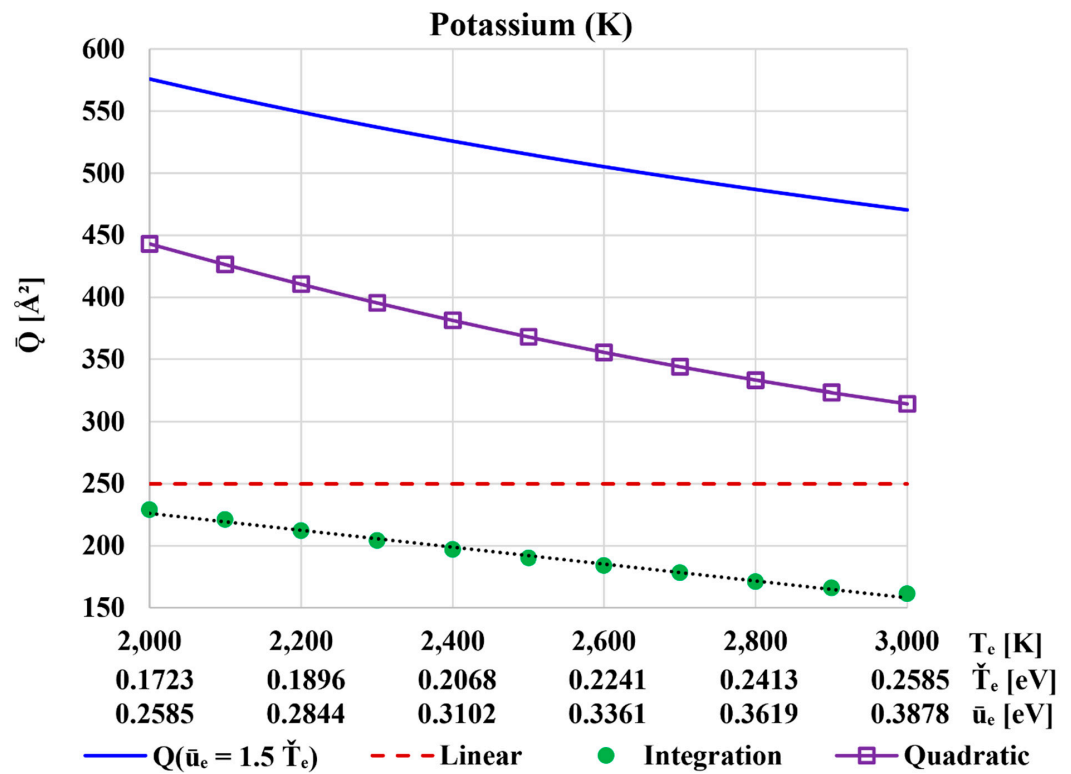

**Figure 11.** Temperature dependence of the average electron–neutral collision cross sections of potassium vapor (K), as obtained using four different methods.

Figure 12 is used to contrast the four predicted curves of the average electron–neutral collision cross sections for molecular oxygen ($O_2$). Methods 1 ($\overline{u}_e = 1.5\check{T}_e$) and 2 (linear) predict a constant value of 5.0326598 Å$^2$ and 3.9 Å$^2$, respectively. Methods 3 (integration) and 4 (quadratic) predict that $\overline{Q}_{O2}$ increases as $T_e$ increases (the opposite behavior to $\overline{Q}_K$ for the three methods that showed a temperature-dependent $\overline{Q}_K$). Based on the arithmetic mean profile of the four individual profiles (including the zero-degree polynomials of methods 1 and 2), the mean rate of change in $\overline{Q}_{O2}$ is 0.0003518 Å$^2$/K. The values of $\overline{Q}_{O2}$ are the smallest among the six gaseous species covered in the present study. As in the case of CO, the $\overline{Q}_{O2}(T_e)$ values in method 4 are described as a linear function (not as a quadratic one).

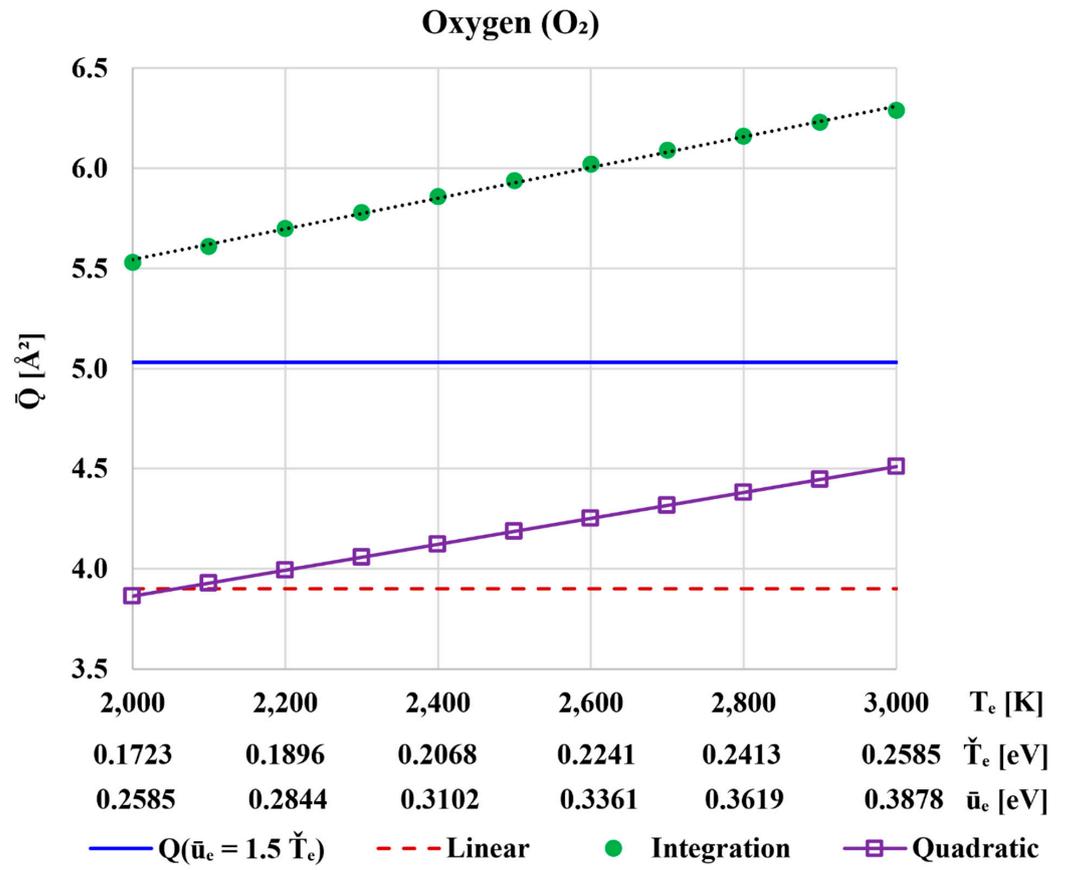

**Figure 12.** Temperature dependence of the average electron–neutral collision cross sections of molecular oxygen ($O_2$), as obtained using four different methods.

### 3.3. Statistical Measure of Disparity in Average Collision Cross Sections

To complement the qualitative estimation of the deviations (or agreement) between the four methods of predicting the average electron–neutral collision cross sections in the previous part (through visually contrasting profile curves), the current part aims to quantitatively summarize the deviations (or agreement) in the predictions for each of the six gaseous species covered herein, and for each of the eleven electron temperatures (between 2000 K and 3000 K) adopted here in the computations and visualizations of the previous part. The selected measure of such deviation is the coefficient of variation (CV or COV) [47], which is the standard deviation divided by the arithmetic mean. This statistical quantity has the advantage of being non-dimensional and also being simple to interpret. It can be safely used here without encountering a singularity because the average electron–neutral collision cross sections (and thus, their arithmetic mean values) are neither zero nor changing their sign but are instead always positive.

The estimated coefficient of variation reported herein is based on the sample-based standard deviation, and it is computed as

$$\text{COV} = \frac{\sqrt{\frac{1}{3}\sum_{i=1}^{4}\left(\overline{Q}_i - \frac{1}{4}\sum_{i=1}^{4}\overline{Q}_i\right)^2}}{\frac{1}{4}\sum_{i=1}^{4}\overline{Q}_i} \times 100\% \qquad (41)$$

where $\overline{Q}_i$ is the predicted average (simplified) electron–neutral collision cross section using the $i$th method.

Table 8 summarizes the estimated coefficient of variation results for the six covered species. For each species, the COV value at each electron temperature is listed, as well as the mean, minimum, and maximum of the eleven COV values of that species. From the mean COV values, the worst deviation among the four ($\overline{Q}$) prediction methods corresponds to potassium (K), with a mean COV of 43.56%, which is the largest of the six species. The least disparity between the four ($\overline{Q}$) predictions corresponds to molecular hydrogen ($H_2$), which has the smallest mean COV of 12.92%. Considering the minimum COV values, it is noted that they do not necessarily correspond to a boundary temperature value (either the lower limit of 2000 K or the upper limit of 3000 K). A similar finding is true for the maximum COV.

**Table 8.** Estimated coefficients of variation (based on the results of 4 prediction methods of $\overline{Q}$) for the average electron–neutral collision cross section for the 6 covered species at the 11 values of electron temperature ($T_e$).

| Index | $T_e$ (K) | Estimated Coefficients of Variation for Each Species ||||||
|---|---|---|---|---|---|---|---|
| | | CO | $CO_2$ | $H_2$ | $H_2O$ | K | $O_2$ |
| 1 | 2000 | 25.95% | 43.10% | 13.03% | 43.94% | 44.13% | 18.19% |
| 2 | 2100 | 26.29% | 43.26% | 13.07% | 42.92% | 43.78% | 18.32% |
| 3 | 2200 | 26.61% | 43.33% | 13.08% | 41.84% | 43.66% | 18.57% |
| 4 | 2300 | 26.89% | 43.24% | 13.06% | 40.70% | 43.53% | 18.78% |
| 5 | 2400 | 27.17% | 43.07% | 13.02% | 39.52% | 43.40% | 19.02% |
| 6 | 2500 | 27.42% | 42.79% | 12.98% | 38.38% | 43.36% | 19.29% |
| 7 | 2600 | 27.72% | 42.32% | 12.92% | 37.33% | 43.28% | 19.59% |
| 8 | 2700 | 27.94% | 41.80% | 12.86% | 36.40% | 43.29% | 19.83% |
| 9 | 2800 | 28.12% | 41.26% | 12.78% | 35.66% | 43.52% | 20.09% |
| 10 | 2900 | 28.31% | 40.65% | 12.68% | 35.04% | 43.54% | 20.37% |
| 11 | 3000 | 28.44% | 39.70% | 12.59% | 34.42% | 43.62% | 20.59% |
| Mean | | 27.35% | 42.23% | 12.92% | 38.74% | 43.56% | 19.33% |
| Minimum | | 25.95% | 39.70% | 12.59% | 34.42% | 43.28% | 18.19% |
| Maximum | | 28.44% | 43.33% | 13.08% | 43.94% | 44.13% | 20.59% |

## 4. Discussion

After the results presented in the previous section, some remarks are given below.

First, the results provided are not enough per se to identify which method among the four methods investigated herein is the best one. This is explained by the absence of a ground truth reference to compare with. However, this does not constitute a deficiency or incompleteness in the presented study, which does not aim to identify or recommend a particular method for computing temperature-dependent electron–neutral collision cross sections. Instead, this study contrasts the predictions of the different four methods, and it both qualitatively and quantitatively assesses the disparity (or coherence) in the predicted results. In the present study, all the investigated methods are considered valid, and none of them is viewed as better than the other. This study emphasizes the presence of large deviations in the computed electron–neutral collision cross sections, which represent just one part of the computation of the plasma electric conductivity. Thus, the findings of the present study justify the large uncertainty in estimating plasma electric conductivity. This also suggests that a simplified modeling approach in computing electron–neutral collision cross sections or plasma electric conductivity can be a sensible choice given that a complicated approach is not guaranteed to provide better estimations.

Second, it was shown that the electron–neutral collision cross sections of carbon monoxide (CO), molecular hydrogen ($H_2$), and molecular oxygen ($O_2$, but according to two methods only out of the four) increase smoothly if the electron temperature increases within the range of temperatures covered (from 2000 K to 3000 K). This indicates more impediment to electrons' mobility at higher temperatures and thus indicates lower plasma electric conductivity. However, this behavior should not misleadingly be interpreted as a higher-temperature alkali-metal seeded plasma that is dominantly composed of one or more of these three species (CO, $H_2$, and $O_2$) having lower electric conductivity than lower-temperature plasma. The reason is that as the plasma temperature increases, the level of thermal ionization (freeing electrons from atoms under the influence of high temperatures) of the seeded alkali-metal element (such as potassium) increases rapidly, following a nonlinear dependence on the temperature [48]. The nonlinear increase in the ionization level (fraction of the seeded alkali-metal atoms that are ionized) with the temperature includes the temperature's exponential term [49,50], such that the strengthening of the electric conductivity due to the improved ionization at elevated temperatures overpowers the opposite effect of weakening the electric conductivity due to the enlarged electron–neutral collision cross sections at elevated temperatures. Thus, the plasma electric conductivity is always boosted at elevated temperatures, even if the average electron–neutral collision cross sections increase with the temperature for one or more gaseous species contained within the plasma gas mixture.

Third, the six gaseous species considered herein for analysis are not intended to be an inclusive list encompassing all species that can possibly appear in the combustion of a wide variety of solid, liquid, and solid fuels. This is neither practical nor intended. Instead, the selection of a few gaseous species that are representative of combustion reactants, combustion products, and seeding alkali metals makes this study more concise and easier to comprehend. The six gaseous species covered herein were selected such that they are all neutral molecular species, without dissociation [51,52], which have data in the four literature sources of the four methods investigated herein. For example, molecular nitrogen ($N_2$) was not included in the current study, because it is not one of the nine gaseous species listed in the literature source of method 3 (integration).

Fourth, the inclusion of molecular oxygen herein (although not an ordinary combustion product gas) is justified, because despite being a reactant oxidizer species that is consumed during combustion, it may still exist in the combustion products (and thus, in the plasma mixture) at a small concentration, in the case of the lean fuel (oxygen-enriched) mode of combustion [53,54]. Similarly, the inclusion of molecular hydrogen (despite being a fuel that is consumed during combustion) in the current study can be justified through the opposite fuel-rich or fuel-enriched combustion modes [55,56], with hydrogen being the fuel or a component of it. In addition, including these two reactant species in the analysis enriched this study by manifesting at least one unique feature in their temperature-dependent electron–neutral collision cross section results, which was not found in the remaining four species covered here.

## 5. Conclusions

Modeling combustion plasma for magnetohydrodynamic (MHD) power generation requires an ability to estimate the electric conductivity of plasma, which, in turn, needs a method to calculate the electron–neutral collision cross section for momentum transfer, taking into account the influence of temperature and the chemical composition of the plasma gas. Motivated by this need, the present study investigated the level of agreement or disagreement between four methods (based on four published studies in the literature) for predicting the electron–neutral collision cross section as a function of electron temperature alone for six relevant gaseous species and for a range of temperatures suitable for both air–fuel combustion and oxy–fuel combustion.

This study showed that using an average electron energy in expressions designed for monoenergetic electron energy gives reasonable results that lie near those obtained using one or more other methods, with the exception of potassium vapor, as this method gives much larger values than any of the other three methods. Also, this study showed that using linear approximations for the collision cross sections seems to be acceptable in the temperature range of interest for combustion, with water vapor being a mild exception. Based on the performed analysis of the coefficients of variation, molecular hydrogen has relatively high consistency in its computed collision cross section, which slowly increases as the electron temperature increases. Carbon dioxide and water vapor (the major combustion products when completely burning a hydrocarbon fuel) become less scattering to electrons as the electron temperature increases, which is another reason (in addition to the intensified thermal ionization) to favor elevated-temperature oxy–fuel-based plasma over air–fuel-based plasma, since the former is associated with higher plasma electric conductivity and thus a better ability to generate electricity.

**Funding:** This research received no external funding.

**Institutional Review Board Statement:** Not applicable.

**Informed Consent Statement:** Not applicable.

**Data Availability Statement:** No dataset was used.

**Conflicts of Interest:** The author declares that this research was conducted in the absence of any commercial or financial relationships that could be construed as potential conflicts of interest.

## Nomenclature

| | |
|---|---|
| $\lambda_e$ | Mean free path of electron (between two successive collisions) (m) |
| $\mu_{e-n}$ | Electron mobility under the influence of electron–neutral collisions (m$^2$/s.V) |
| $\nu_{e-n}$ | Electron–neutral collision cyclic frequency (Hz, 1/s) |
| $\bar{\nu}_{e-n}$ | Average electron–neutral collision cyclic frequency (Hz, 1/s) |
| $\sigma_{e-n}$ | Electric conductivity of plasma under the influence of electron–neutral collisions (S/m) |
| $\tau_{e-n}$ | Time between two successive electron–neutral collisions (s) |
| $\bar{\tau}_{e-n}$ | Average time between two successive electron–neutral collisions (s) |
| $e$ | Electron charge magnitude, $1.602176634 \times 10^{-19}$ (C) [57] |
| eV | Electronvolt or electron volt, $1.602176634 \times 10^{-19}$ (J) [58] |
| $k$ | Boltzmann constant, $1.380649 \times 10^{-23}$ (J/K) [59] |
| $\hat{k}$ | Transformed Boltzmann constant, $k/e = 8.617333262 \times 10^{-5}$ (eV/K) [60] |
| $m_e$ | Electron mass, $9.1093837015 \times 10^{-31}$ (kg) [61] |
| $n_e$ | Number density of electrons as detached particles in a plasma gas (1/m$^3$) |
| $n_s$ | Number density of neutral particles of a generic gaseous species labeled "s" (1/m$^3$) |
| $Q$ | Monoenergetic (dependent on the electron speed or the electron kinetic energy) electron–neutral collision cross section (Å$^2$) |
| $\overline{Q}$ | Average (simplified, independent of electron speed or kinetic energy) electron–neutral collision cross section (Å$^2$) |
| $T_e$ | Electron absolute temperature (K) |
| $\check{T}_e$ | Transformed electron absolute temperature expressed as energy, $\hat{k}_B T_e$ (eV) |
| $u_e$ | Monoenergetic kinetic energy of electron (eV) |
| $\bar{u}_e$ | Average kinetic energy of electron (eV) |
| $v_e$ | Monoenergetic electron thermal speed (m/s) |
| $\bar{v}_e$ | Average electron thermal speed, based on Maxwell–Boltzmann distribution (m/s) |